\documentclass[fleqn,usenatbib]{mnras}

\usepackage{newtxtext,newtxmath}

\usepackage[T1]{fontenc}

\DeclareRobustCommand{\VAN}[3]{#2}
\let\VANthebibliography\thebibliography
\def\thebibliography{\DeclareRobustCommand{\VAN}[3]{##3}\VANthebibliography}

\usepackage{graphicx}	
\usepackage{amsmath}	

\title[One-d Shocks with PIC]{Novel Methods for Simulating Astrophysical Plasmas and the Coherent Emission in Fast Radio Bursts}

\author[Lloyd-Ronning et al.]{
Nicole M. Lloyd-Ronning,$^{1,2}$\thanks{E-mail: lloyd-ronning@lanl.gov}, Patrick Kilian,$^{3}$
Guangye Chen,$^{4}$ 
\newauthor \ Chengkun Huang$^{4}$, Fan Guo$^{5}$, Lucian Sahd$^{4}$, Makana Silva$^{1,2}$
\\
$^{1}$Computational Physics and Methods Group, Los Alamos National Lab, Los Alamos, NM 87544\\
$^{2}$Center for Theoretical Astrophysics, Los Alamos National Lab, Los Alamos, NM 87544\\
$^{3}$Space Sciences Institute, Boulder, CO, 80301\\
$^{4}$Applied Mathematics and Plasma Physics Group, Los Alamos National Lab, Los Alamos, NM 87544\\
$^{5}$Nuclear Particle Physics, Astrophysics and Cosmology Group, Los Alamos National Lab, Los Alamos, NM 87544\\
}

\date{Accepted XXX. Received YYY; in original form ZZZ}

\pubyear{2024}

\begin{document}
\label{firstpage}
\pagerange{\pageref{firstpage}--\pageref{lastpage}}
\maketitle

\begin{abstract}
 We present particle-in-cell simulations of one dimensional relativistic electromagnetic shocks in a uniform magnetic field, for a range of magnetic field strengths, plasma temperatures and numerical initial conditions.  We show that the particle energy distributions of these shocks can develop a state of population inversion in the precursor and shock regions, which may allow for synchrotron maser (or maser-like, coherent) emission.  Our set-up is applicable to conditions expected in models of fast radio bursts and therefore lends credence to the synchrotron maser model for these transients.  We also show, for the first time, how a newly developed ``analytic particle pusher'' for kinetic simulations gives similar results to the commonly-used Boris pusher, but for larger timesteps and without the need to resolve the gyro-radius and gyro-period of the system.  This has important implications for modeling astrophysical plasmas in extreme magnetic fields as well as for bridging scales between kinetic and fluid regimes.  
\end{abstract}




\section{Introduction}



 Many astrophysical systems involve highly magnetized collisionless plasmas (where the mean free path for collisions is much larger than the scale of the system); these systems can also be highly relativistic, with the bulk flow velocity and/or the internal velocities of the particles nearing the speed of light.  Examples include relativistic jets in active galactic nuclei \citep{Mar06, Rom17, Bland19}, gamma-ray burst jets \citep{Pir04, Kum15}, pulsar wind nebulae \citep{Gaens06}, and fast radio bursts \citep{Pet22}.  Because an analytical treatment of the plasmas in these problems is often untenable, capturing the correct plasma physics interactions requires high fidelity kinetic modeling, usually with particle-in-cell (PIC) codes \cite[see, for example, the recent review by][and references therein]{Nish21}. However, the collisional mean free path and the characteristic length scales of the system are often many orders of magnitude larger than the gyro-radius, so accurately modeling the physics of these systems over observable timescales can be computationally too expensive; fluid approximations and hybrid techniques must be employed, at the expense of losing some of the important kinetic plasma physics.  
  \\

 A Fast Radio Burst (FRB), with its millisecond timescale bright and coherent radio emission, offers a unique laboratory to study plasma physics under extreme conditions of high magnetic fields and relativistic velocities.  There is some evidence that these transients are associated with magnetars \citep{Chime20, Boch20, Hu24}, neutron stars with magnetic fields on the order of $\sim 10^{15}$ Gauss.  There exists a range of models for what might produce such short timescale, coherent radio emission including synchrotron maser emission, curvature emission of electron bunches, reconnection of current sheets, coherent emission from a relativistic shock front, and more \citep[for recent reviews see, e.g., ][]{Pet19, CC19, Platt19,Lyu21, Zhang23}. \\

Several studies \citep{Lyu14, Mur16, Bel17, Wax17, Ghi17, Metz19, PS19, Bab20, Marg20a, Marg20b, Sir21} have suggested that the synchrotron maser model is a viable emission mechanism for FRBs (particularly within the magnetar scenario for their progenitor systems).   A maser can occur when there exists a resonantly unstable system that causes a negative total absorption coefficient, or population inversion of radiating particles \citep{AA88, HA91}. Masers have been observed in a number of astrophysical systems for decades \citep[e.g.][]{Weav65,Cheung69,Ball70,Snyd74}; for a review, see \cite{Elitz92}. \\

\cite{Lyu21} discuss some key requirements and properties of synchrotron maser emission in particular. For example, they show that the growth rate of the instability is $\kappa \sim 0.3 \sigma^{1/3} \omega_{c}$, for $\sigma > 1$, where $\sigma$ is the magnetization parameter and $\omega_{c}$ is the cyclotron frequency. The characteristic frequency of the amplified wave is $\omega \sim \sigma^{1/2}\omega_{p}$, where $\omega_{p}$ is the plasma frequency. The formation of the maser is roughly temperature independent as long as $T< 0.03 m_{e}c^{2}$, where $m_{e}$ is the mass of the electron and $c$ is the speed of light \citep[see also][]{AA06}.  Finally, the width of the unstable region is $\sim 2c \omega_{c}^{-1}$; written in terms of the plasma inertial length, $d_{e} = c \omega_{p}^{-1}$, this width of the unstable region is $\sim 2 d_{e}/\sqrt{\sigma}$.   

For this type of coherent emission, there is a characteristic ring-like formation in momentum space in addition to the inversion in the particle energy spectrum. According to \cite{Lyu21}, this ring-like particle distribution can form as a natural consequence of a collisionless shock in a magnetized flow that is mediated by Larmor rotation. When the upstream flow enters the shock, the particles are slowed substantially and begin to rotate in the magnetic field (which is enhanced at the shock front). Finally, we note that the emission frequency of the maser does not significantly exceed the plasma frequency: the characteristic frequency of the emission is $\omega_{{\rm maser}} \sim \upzeta \omega_{p}$, where $\upzeta \sim$ few.\\

Relativistic magnetized perpendicular shocks can create semi-coherent electromagnetic waves that travel just ahead of the shock front, the so-called precursor wave. A population inversion can occur at the shock front where the incoming particles rotate around the shock compressed magnetic field, which can then lead to coherent synchroton maser emission. PIC simulations of relativistic magnetized electron-positron pair shocks have been performed by a number of groups \citep{Lang88, Gall92, SS09, Iwa17, Iwa18, Plot18, PS19}.  In particular, \cite{PS19} present 1-D simulations \cite[and subsequent 2-D and 3-D simulations that confirm their 1-D results in][]{Bab20, Sir21} of a relativistic shock in a uniform background magnetic field, focusing on the model of synchrotron maser emission from FRBs. They assert that the precursor wave of the magnetic field creates a population of particles that is stably sustained at higher energy levels, characterized by higher transverse momenta (perpendicular to the shock flow and direction of the background field). The shock structure displays a soliton-like feature (a sharp spike in the density), in which particles gyrate around the compressed magnetic field.  This leads to a ``ring" in momentum phase space that can indicate \citep{Lyu06, Lyu21} a negative absorption coefficient, and conditions conducive to a maser. \\

In this paper, we present PIC simulations of one dimensional relativistic shocks embedded in a uniform magnetic field, similar to the simulations in \cite{PS19} but extended further, exploring a wider range of initial particle temperatures and background magnetic field strengths.  We show that a cavity or population inversion develops in the particle energy spectrum at high energies, and argue this can support a synchrotron maser model for FRB emission. Furthermore,  we have developed a novel implementation of a particle pusher in the PIC algorithm that allows us to circumvent the need to resolve the gyro-radius \citep[][and G. Chen et al. 2025]{Chen13, Chac17}. The idea of this pusher relies on transforming to a frame where the electric and magnetic field are parallel, in which an exact solution to the equations of motion may be found.  The method is applicable over scales for which the magnetic field is approximately uniform, but does {\em not} require that the gyro-radius be resolved. Motivated by the need to reduce the computational cost of PIC simulations in high magnetic fields (where the gyro-radius needs to be resolved), we show results of simulations of our 1-D relativistic shock set-up using this novel method for the particle pusher in the PIC simulation.\\

Our paper is organizes as follows: In \S 2 we describe our simulation set-up. In \S 3, we present results of our one dimensional shock simulations over a range of magnetic fields and temperatures, and show how in some cases a cavity-like feature develops in the particle energy spectrum in the region right at the shock front.  We show results using both the Boris pusher --- in which the gyro-radius needs to be highly resolved --- versus the analytic pusher, which does not have this constraint.  Finally, we present our discussion and conclusions in \S 4.

\begin{figure*}
\hspace{-1.0cm}
 \includegraphics[width=16cm]{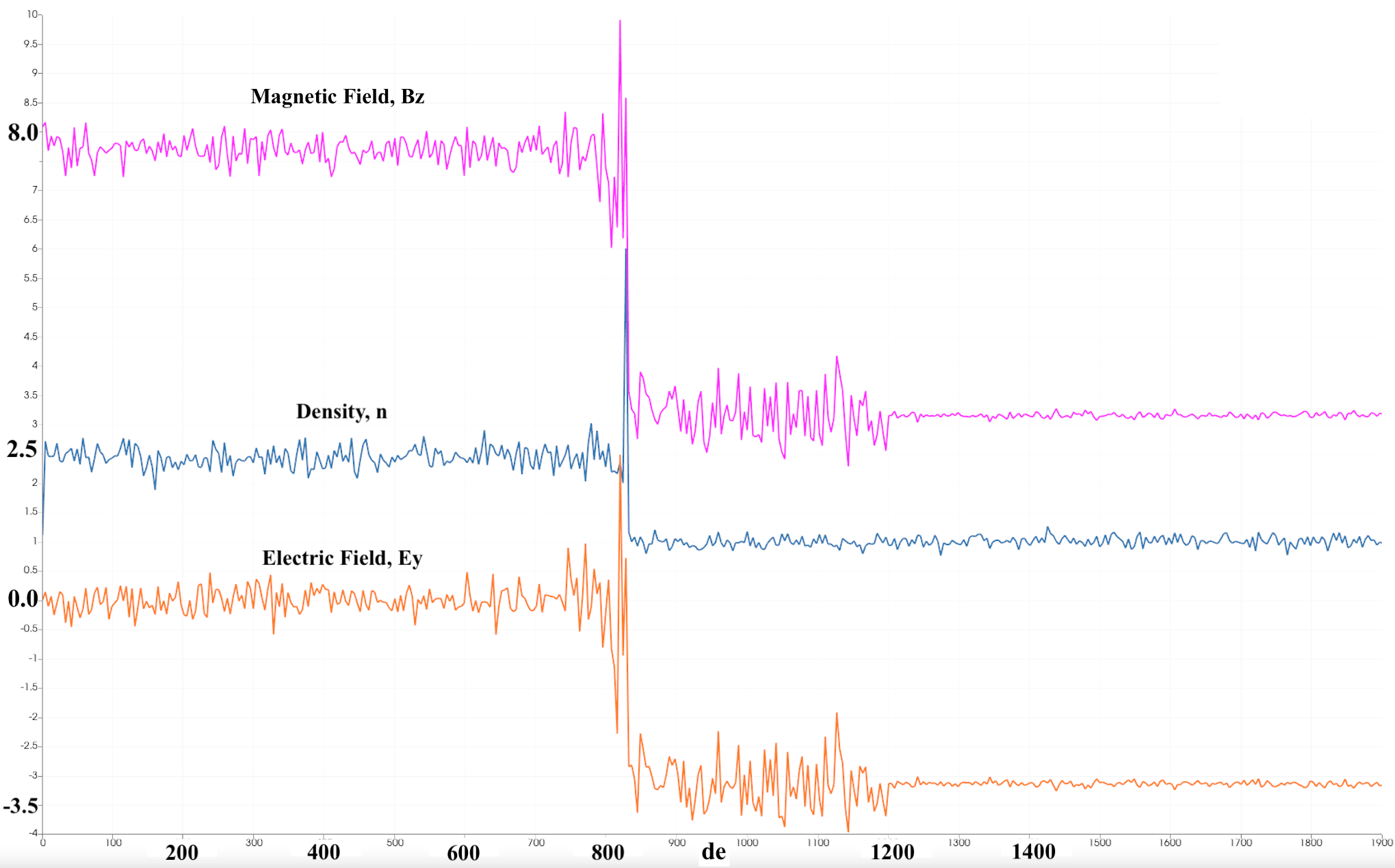} 
    \caption{Normalized magnetic field in the flow direction (magenta), density (blue) and electric field in the perpendicular direction (orange) as a function of position along the flow direction in units of inertial length for a one-D shock with $\Gamma=10$, a temperature of $T=0.0001 m_{e}c^{2}$, and an upstream magnetic field of $B_{\rm PIC}=3.1$ in PIC units, corresponding to a magnetic energy to rest mass energy ratio of $\sigma=10$. Note the precursor wave that develops.}
    \label{fig:nbe}
\end{figure*}

\begin{figure*}
	\includegraphics[width=12.2cm, height=8.5cm]{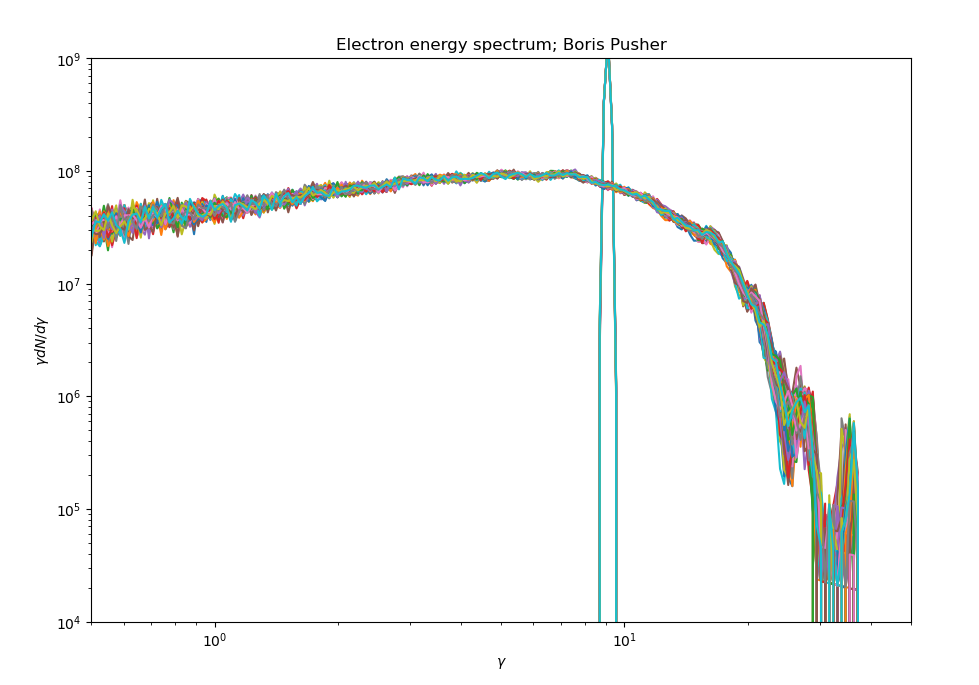}\\
 \includegraphics[width=7.9cm, height=5.6cm]{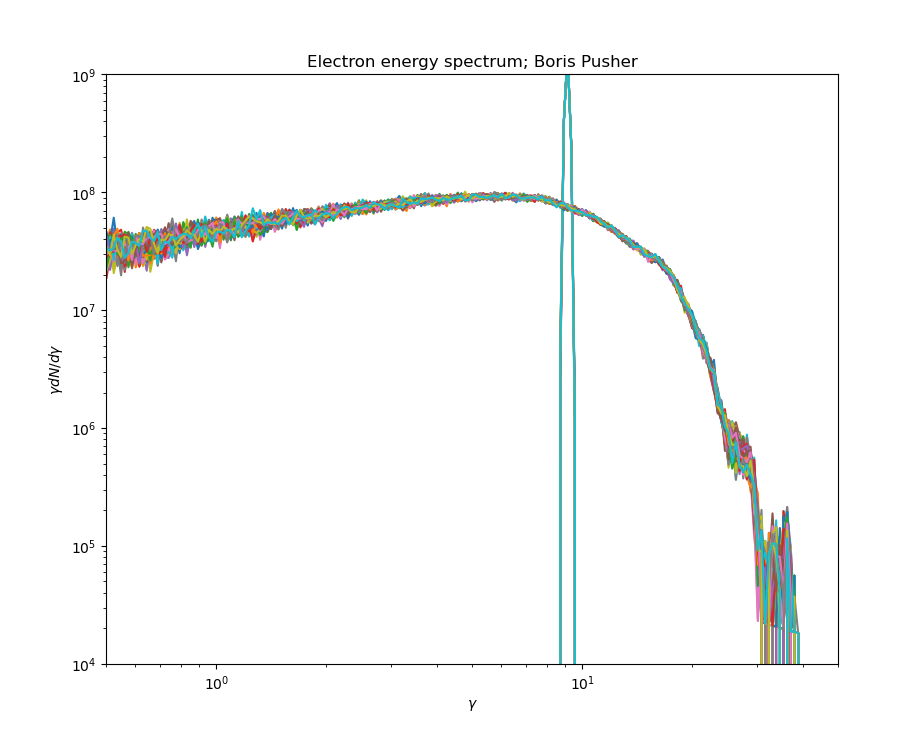} 
 \includegraphics[width=7.7cm, height=5.5cm]{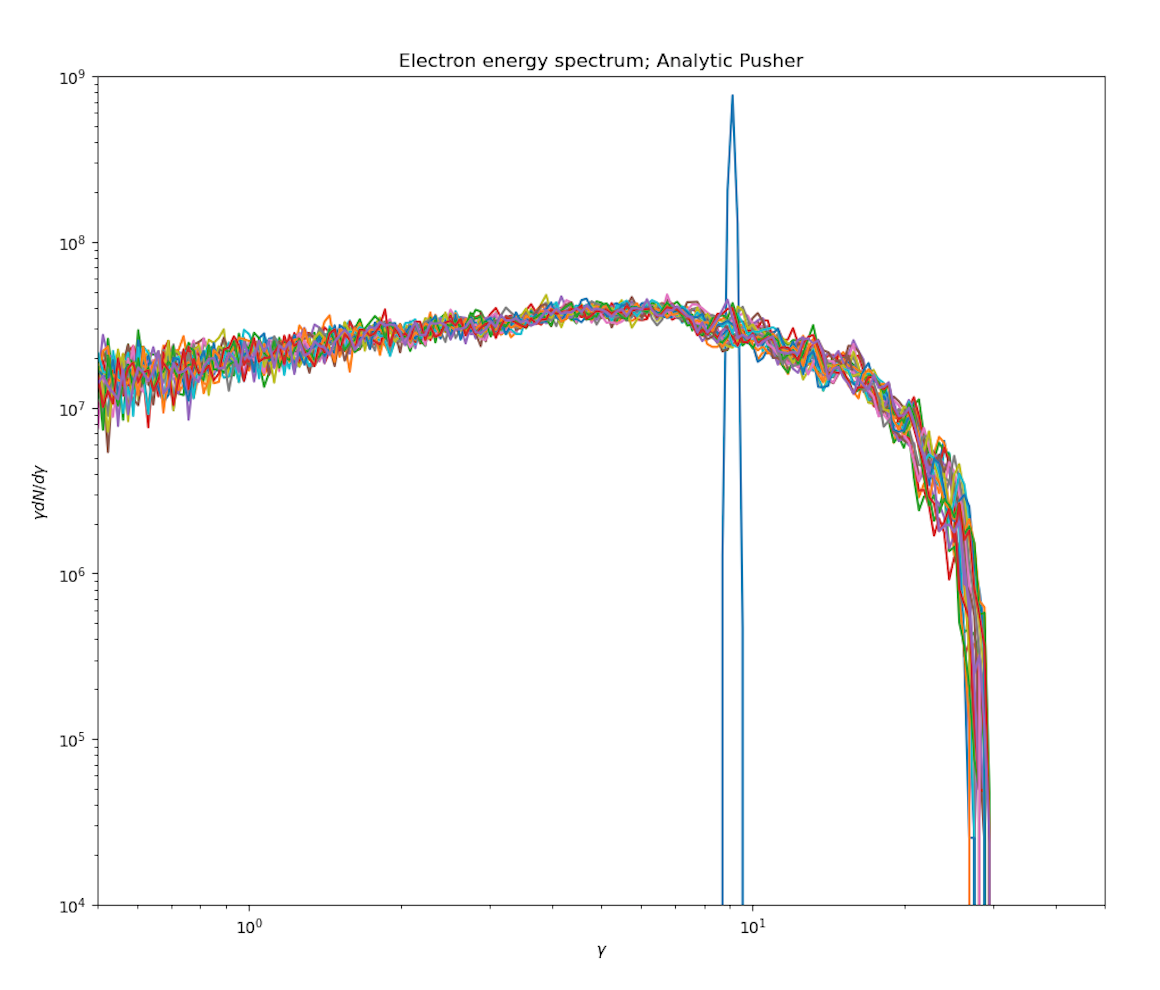}
    \caption{{\bf Top Panel:} Electron energy spectrum in a narrow region in front of the shock  (at $\Delta x=5 {\rm d}_{e}$) for a one dimensional shock with a bulk Lorentz factor of $\Gamma=10$, an initial electron temperature of $T=0.0001 m_{e}c^{2}$, and a PIC magnetic field of $B_{\rm PIC}=3.1$, corresponding to a magnetic energy to rest mass energy ratio of $\sigma=10$. {\bf Bottom Panels:} The left panel shows the same region at later times as the shock has just passed.  The right panel shows the electron energy spectrum far downstream from the shock front. }
    \label{fig:elspecsig10}
\end{figure*}

\begin{figure*}
	\includegraphics[width=12cm]{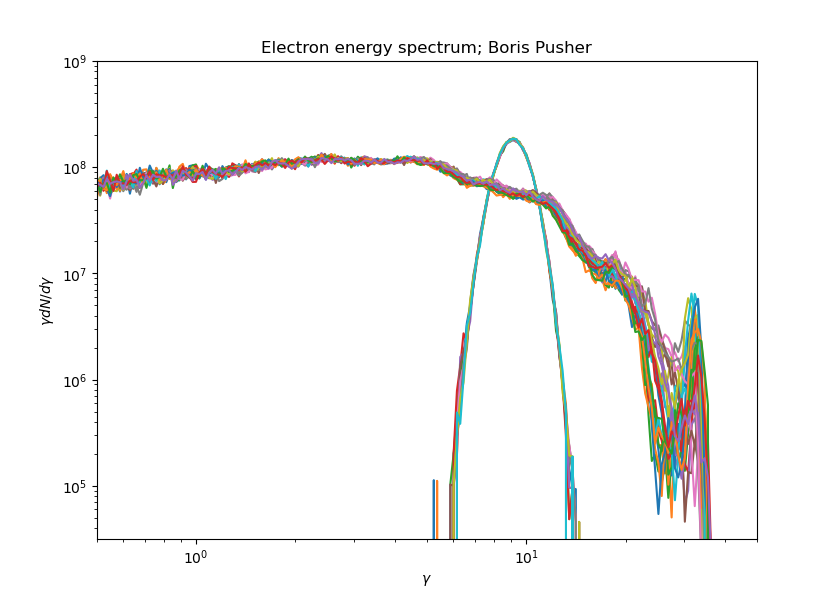} 
    \caption{Electron energy spectrum in a narrow region at the shock front for a one dimensional shock with a bulk Lorentz factor of $\Gamma=10$ an initial electron temperature of $T=0.01 m_{e}c^{2}$, and a PIC magnetic field of $B_{\rm PIC}=3.1$, corresponding to a magnetic energy to rest mass energy ratio of $\sigma=10$.}
    \label{fig:elspecsig10b}
\end{figure*}

\section{Simulation Set Up}
 We follow the set-up of \cite{PS19} (hereafter PS19), who simulate a 1d3v (one spatial dimension, but three dimensions in velocity space) collisionless shock consisting of an electron-positron pair plasma, with a bulk Lorentz factor of $\Gamma=10$, a perpendicular background magnetic field $\vec{B}_{o} = B_{o}\hat{z}$, and the corresponding electric field $\vec{E}_{o} = E_{o}\hat{y} = -\beta_{o}B_{o} \hat{y}$, where $\beta_{o}$ is the initial flow velocity (corresponding to the Lorentz factor $\Gamma$) in units of the speed of light.  The initial temperature of the pairs in their plasma is $T=.0001 m_{e}c^{2}$, and they run their simulations over a range of background magnetic fields ranging from $\sigma=0.3$ to $1000$, where $\sigma$ is the magnetization parameter defined as the ratio of magnetic field energy density to rest mass energy density.   They use a time-step resolution $\Delta t \omega_{p} = 0.005$.    \\

 We initially run exactly the same set up  as PS19 as a check of the validity of our simulations.  We employ their initial time step, by enforcing the so-called magic time-step where $\Delta t = hx/2c$, and $hx = Lx/nx = 4096/409600$, where $Lx$ is the box length along the x-direction and $nx$ is the number of cells in that direction.  
 We run our simulations over a wider range of initial temperatures ranging from $T=.0001 m_{e}c^{2}$ up to $T=0.7 m_{e}c^{2}$ (we note that \cite{Bab20} ran their 2-D simulations over a wider range of temperatures than in PS19 and find similar results to what we find below).  We explore a set of uniform magnetic field values oriented perpendicular to the direction of flow, with values of the normalized field ranging from $B_{\rm PIC} = 0.1$ to $30$, corresponding roughly to the range of $\sigma$ values explored in PS19 (see below for definition of the magnetic field in PIC units and how that corresponds to the magnetization parameter).  \\
 
 We then run simulations with increased time-step/decreased grid resolution and demonstrate how our analytic particle pusher produces similar results to those seen in the more resolved simulations employing the Boris pusher. Note that the gyro-radius in our simulations ranges from $\sim 0.001 d_{e}$ up to $\sim 0.03 d_{e}$ depending on temperature and strength of the magnetic field, while our cell size is $0.01 d_{e}$.

\subsection{Physical vs Simulation Plasma Parameters}
When interpreting our results, it is important to note the relation between scaled PIC units and physical plasma parameters.  

\begin{itemize} 
\item The physical plasma frequency is defined as $\omega_{p} = (\pi n q^{2}/\gamma_{e} m_{e})^{1/2}$, where $q$ is the electron charge, $n$ is the plasma number density, $m_{e}$ is the mass of the electron,  and $\gamma_{e}$ is the internal electron energy Lorentz factor (not the bulk flow Lorentz factor $\Gamma$). \\

\item The relativistic gyro-frequency is $\omega_{c} = qB/(\gamma_{e} m_{e} c)$, where $c$ is the speed of light.\\

\item In our PIC simulations, we use a normalized magnetic field  $B_{\rm PIC} = B*(q/\omega_{p}m_{e})$. When comparing to simulations that use the magnetization parameter $\sigma$, one can use the following relation: $\sigma = (\omega_{c}/\omega_{p})^{2} = (B_{\rm PIC}/\gamma_{e})^{2}$.\\

\item The plasma inertial length is defined by $d_{e} = \omega_{p}/c$.\\

\item The gyro-radius is given by $r_{g} =  \beta_{\perp} d_{e}/B_{\rm PIC}$, where $\beta_{\perp}$ is $v_{\perp}/c \propto v_{{\rm thermal}} \propto T^{1/2}$, and $v_{{\rm thermal}}$ is the thermal velocity of the particles and $T$ the temperature of the electrons and positrons.\\

\item The Debye length is $\sim  d_{e}T^{1/2} \sim (T/\gamma_{e})^{1/2}$, again where $\gamma_{e}$ corresponds to the internal electron energy Lorentz factor, {\em not} the bulk flow Lorentz factor $\Gamma$. \\

\end{itemize}

\noindent Our PIC simulations employ units in which $\omega_{p} = c = 1$.  PIC length scales are normalized to the inertial length $d_{e}$ and timescales to the inverse plasma frequency $\omega_{p}^{-1}$.  \\

Below, we present the outcome of our simulations.

\section{Results}

Similar to PS19, we find a precursor wave ahead of the shock front in the electric and magnetic fields, as seen in Figure~\ref{fig:nbe}.  In their paper, PS19 suggest that in this region and at the shock front, there is a separation of the particles in momentum phase space that is suggestive of a maser instability.  We investigate this claim through a thorough analysis of the particle energy spectrum. \\

\begin{figure*}
	\includegraphics[width=12cm]{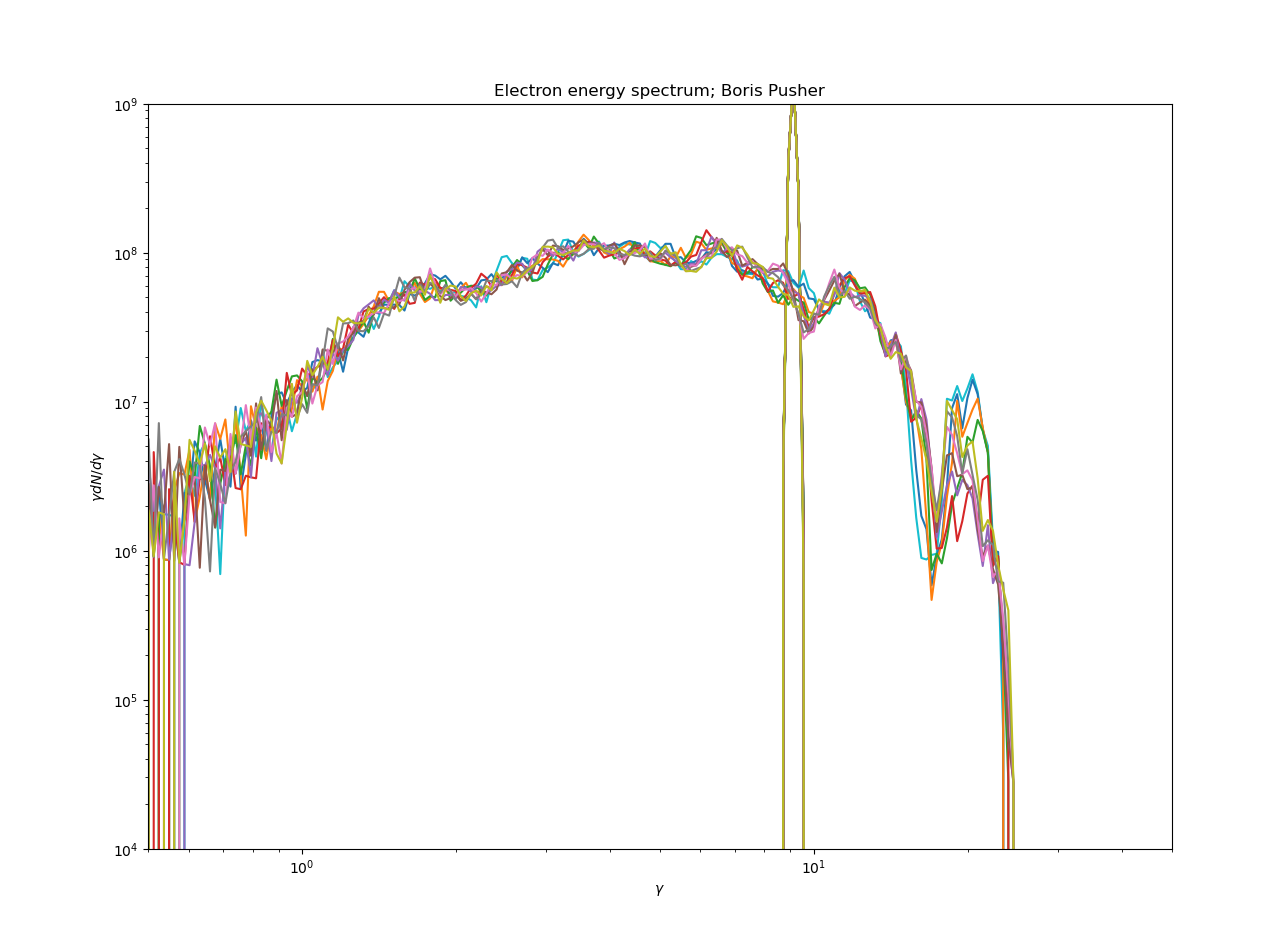} 
    \caption{Electron energy spectrum in a narrow region at the shock front for a one dimensional shock with a bulk Lorentz factor of $\Gamma=10$, an initial electron temperature of $T=0.0001 m_{e}c^{2}$,  and a PIC magnetic field of $B_{\rm PIC}=5.5$, corresponding to a magnetic energy to rest mass energy ratio of $\sigma=30$.}
    \label{fig:elspecsig30}
\end{figure*}

\begin{figure*}
	\includegraphics[width=8.5cm]{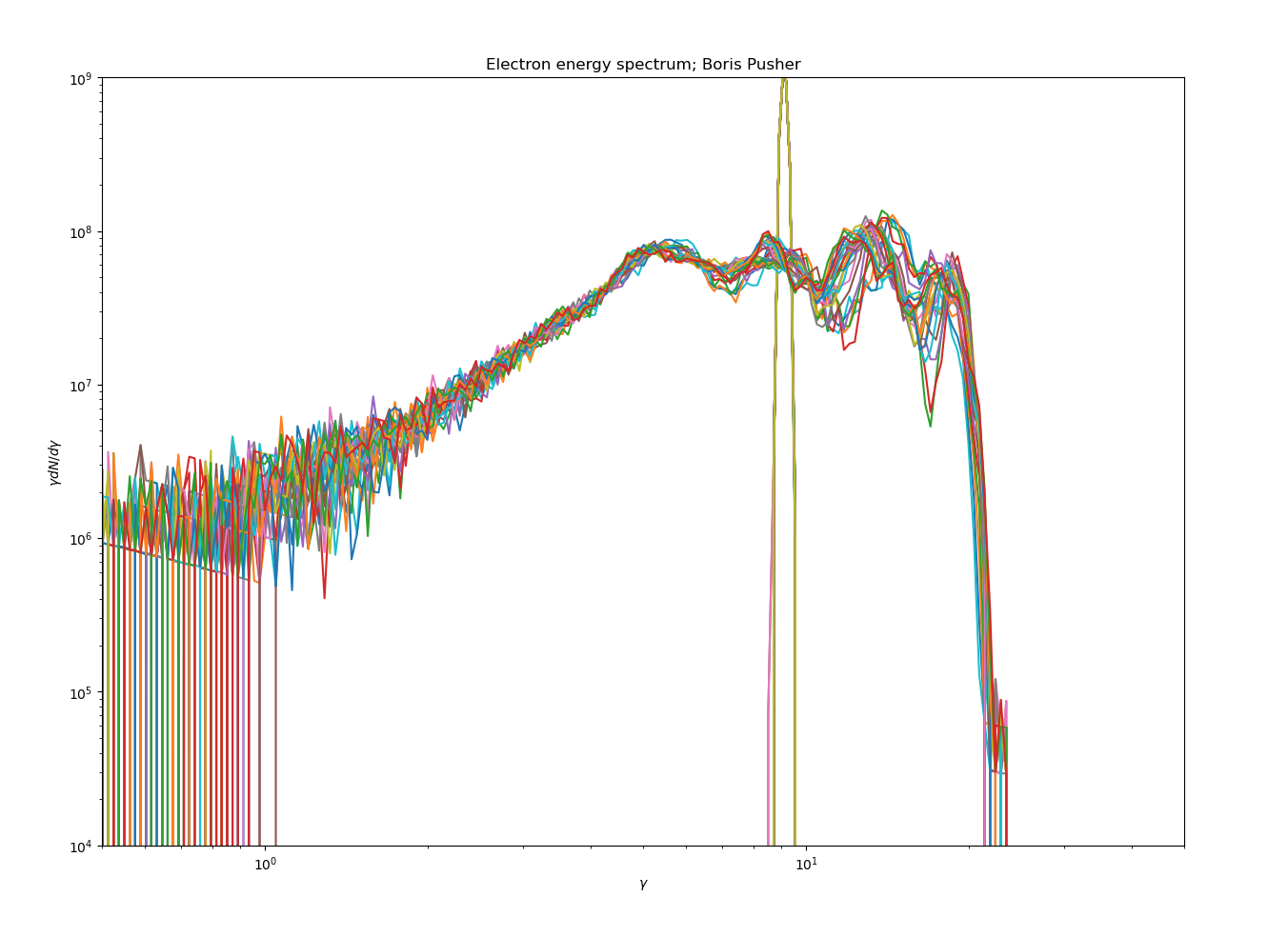} \includegraphics[width=8.5cm, height=6.2cm]{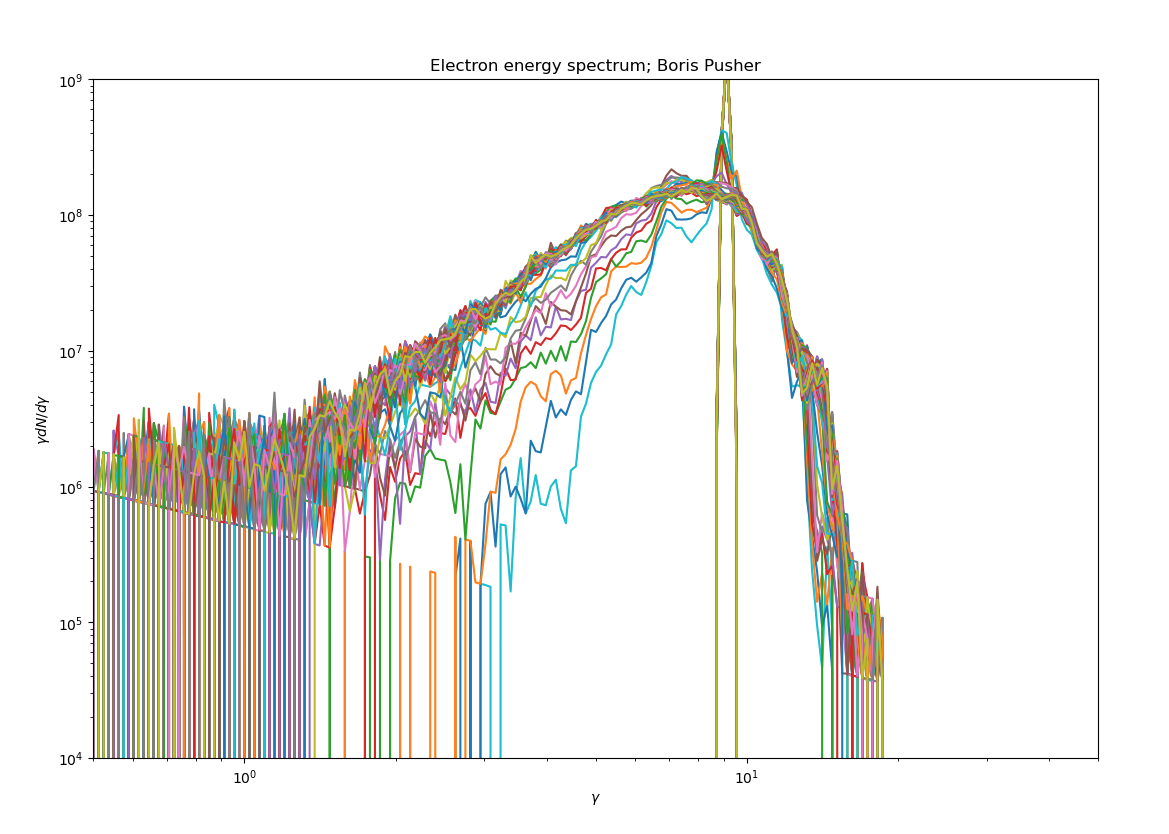}
    \caption{{\bf Left Panel:} Electron energy spectrum in a narrow region at the shock front for a one dimensional shock with a bulk Lorentz factor of $\Gamma=10$ and a PIC magnetic field of $B_{\rm PIC}=17$, corresponding to a magnetic energy to rest mass energy ratio of $\sigma= 289$.{\bf Right Panel:}Electron energy spectrum in a narrow region at the shock front for a one dimensional shock with a bulk Lorentz factor of $\Gamma=10$ and a PIC magnetic field of $B_{\rm PIC}=30$, corresponding to a magnetic energy to rest mass energy ratio of $\sigma=900$. }
    \label{fig:elspecsig900}
\end{figure*}

\begin{figure*}
	\includegraphics[width=7.9cm, height=6.7cm]{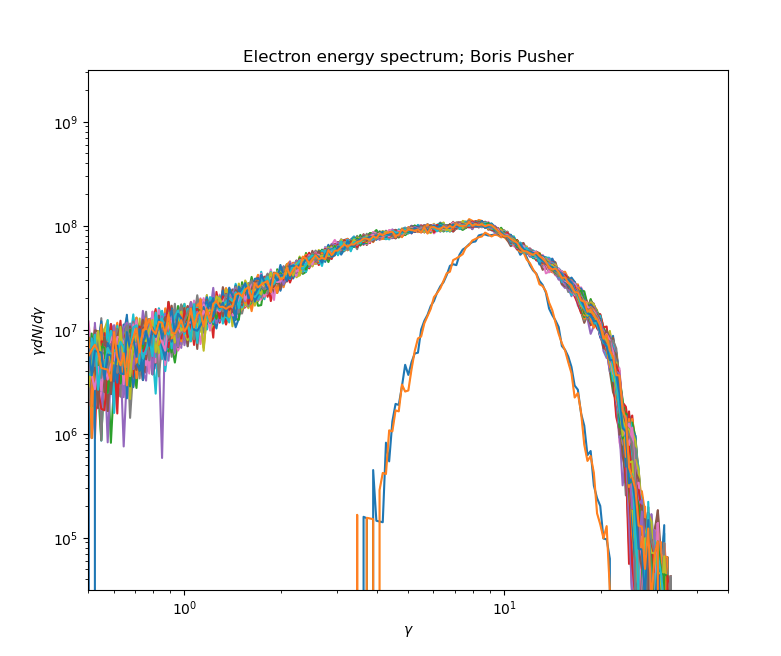} \includegraphics[width=7.9cm, height=6.7cm]{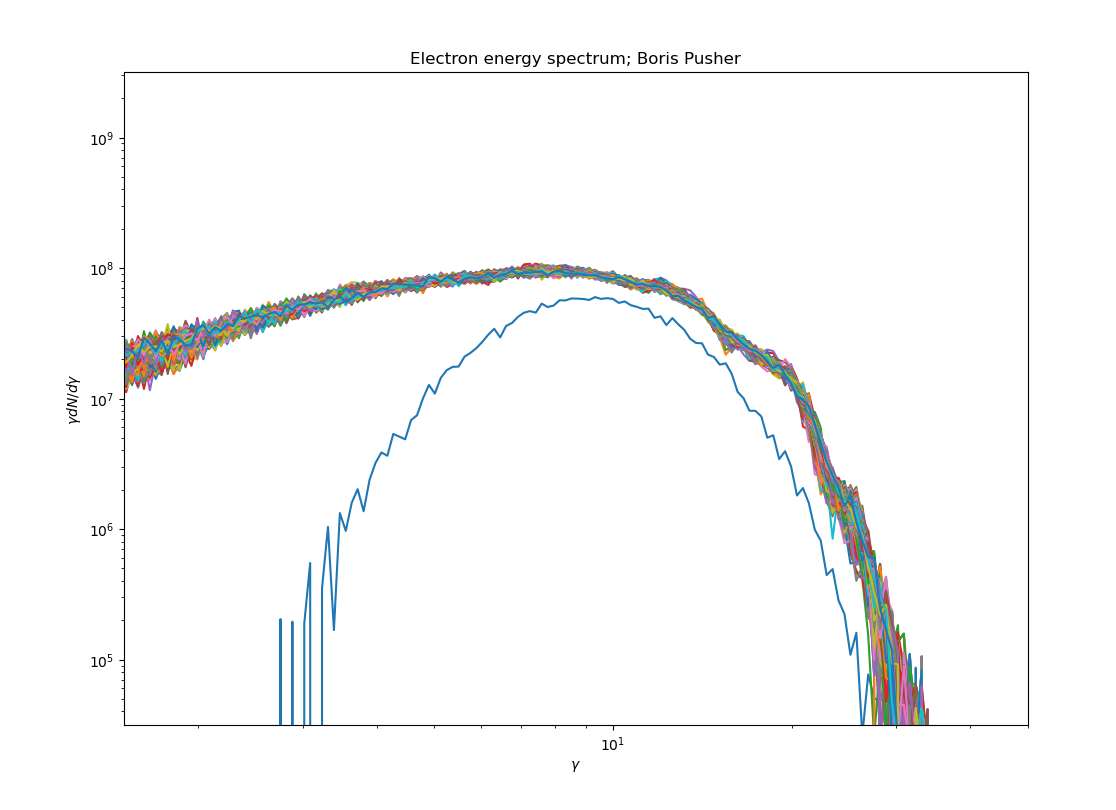} \\
 \includegraphics[width=7.9cm, height=6.7cm]{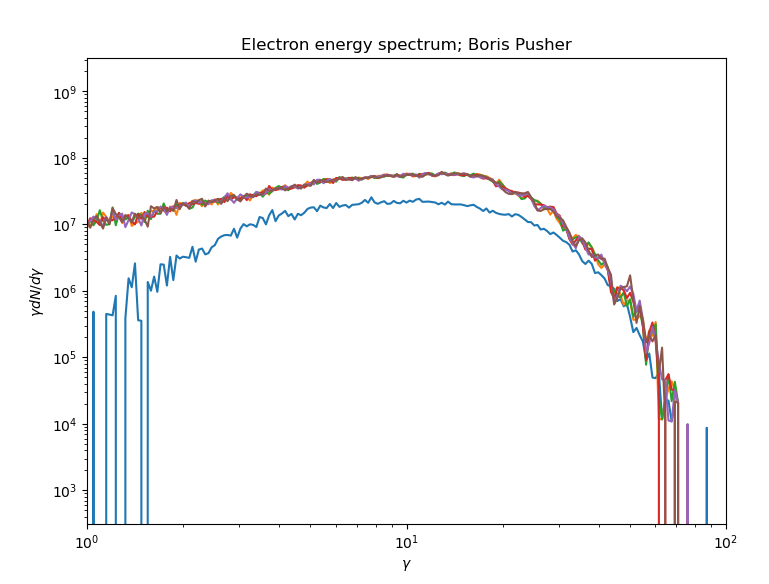}
    \caption{One dimensional shock with a bulk Lorentz factor of $\Gamma=10$, a PIC magnetic field of $B_{\rm PIC}=3.1$, corresponding to a magnetic energy to rest mass energy ratio of $\sigma=10$, and an initial electron temperature of $T=0.05 m_{e}c^{2}$ (top left), $T=0.1 m_{e}c^{2}$ (top right), and $T=0.7 m_{e}c^{2}$ (bottom). }
    \label{fig:elspecsigTemp}
\end{figure*}

Figure~\ref{fig:elspecsig10} - Figure~\ref{fig:elspecsigTemp} show the electron energy spectra over a range of initial conditions, with each curve in the plot corresponding to a different time slice. We focus on a few of the most relevant cases from our wide range of parameter space explored. For magnetic fields ranging from $B_{\rm PIC} \approx 0.1 - 20$, the particle energy spectra show a sustained feature at the point of the precursor and shock front - a dip and then rise in the particle energy spectrum - which we interpret as a population inversion that could lead to synchrotron maser emission. We note that a requirement for instability is roughly that $df/d\gamma > f/\gamma$ in the particle energy distribution - i.e. the slope of the particle energy distribution needs to be sufficiently steep to satisfy the requirements of instability.  This condition is indeed met in our particle energy distributions. \\

Figure~\ref{fig:elspecsig10} shows our results in the narrow region around the pre-cursor and shock front (top panel) for the PS19 set-up described in \S 2 (i.e. $\Gamma = 10, \ T=.0001 m_{e}c^{2}$, and the same numerical resolution used in their paper), for a magnetization parameter of $\sigma=10$.  The bottom panels of this figure show the particle energy spectra at later times in this region just as the shock has passed (left panel) and then far downstream (right panel), well after the shock has passed. Figure~\ref{fig:elspecsig10b} shows the same as Figure~\ref{fig:elspecsig10} but for a temperature of $T=0.01m_{e}c^{2}$, while Figure~\ref{fig:elspecsig30} shows the same as Figure~\ref{fig:elspecsig10} but for a magnetization parameter of $\sigma=30$.  As we increase the magnetization parameter, this ``dip'' feature in the high energy part of the spectrum appears to disappear, as seen in Figure~\ref{fig:elspecsig900} for $\sigma=289$ (left panel) and $\sigma = 900$ (right panel). \\

\begin{figure}
 \includegraphics[width=7.9cm, height=6.7cm]{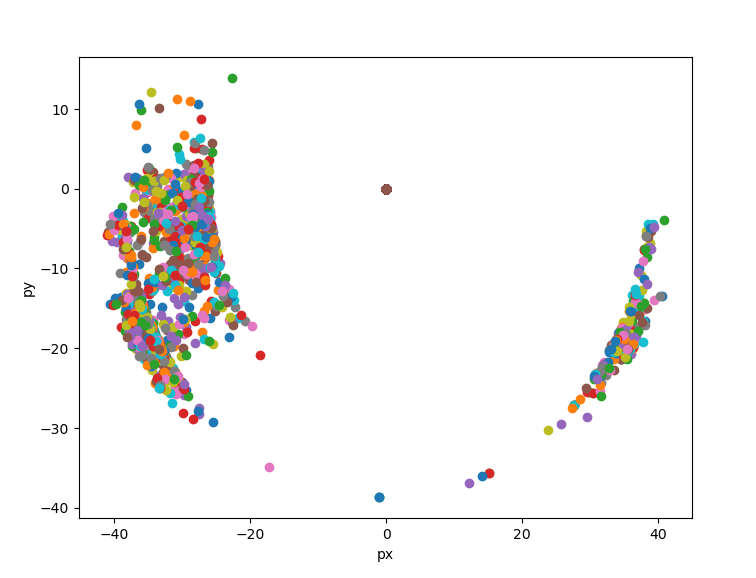} 
    \caption{Momentum (normalized to the initial thermal velocity/momentum) in the y direction vs. momentum in the x-direction (the background magentic field, with $\sigma=10$ is in the z-direction). }
    \label{fig:momentum}
\end{figure}

As discussed in \cite{Lyu21}, the population inversion should disappear in this type of physical set-up for temperatures above about $0.03 m_{e}c^{2}$, and indeed we see that this is the case in three panels of Figure~\ref{fig:elspecsigTemp} which show the particle energy spectrum for $T=0.05 m_{e}c^{2}$ (top left panel), $T=0.1 m_{e}c^{2}$ (top right panel), and $T=0.7 m_{e}c^{2}$ (bottom panel), for a magnetization parameter $\sigma=10$. Similar results were found in the 2-D simulations of \cite{Bab20}.  Finally, as mentioned above, one of the signatures of maser-like emission \citep{Lyu21} is a ring-like feature that develops in momentum space.  We see this ring developing, presented in Figure~\ref{fig:momentum}. \\





\subsection{Demonstration of the Analytic Pusher}

One of the goals of this paper is to show the utility of a newly developed analytic particle pusher for implementation in PIC codes (Chen et al. 2025). When the electric and magnetic fields are uniform, it is almost always possible to perform a Lorentz transformation to a reference frame in which the electric and magnetic fields are parallel. In this case, there exists an exact analytic solution for the particle trajectories. In the more general case where the fields are non-uniform, one is limited by the length scale over which the fields are approximately uniform.  For high magnetic fields, this is often much larger than the gyro-radius of the system. Our analytic particle pusher is implemented in a time-centered fashion, similar to leap-frog or mid-point schemes, so that it is second-order accurate. \\ 

To advance particles, we solve the following equations of motion (EOM) in the lab frame:
\begin{equation}
    d\vec{u}/d\tau=F \vec{u}, d\vec{X}/d\tau = \vec{u}, 
\end{equation}
\noindent where $\tau = t/\gamma$ is proper time, t is time in the lab frame, $\gamma$ is the Lorentz factor
$\vec{u} = (\gamma c, \gamma \vec{v})$  and $\vec{X}=(ct,\vec{x})$ are the four-velocity and four-position with $\vec{v}$ and $\vec{x}$ the particle velocity and position respectively, c is the speed of light, and 
$F$ is the Faraday tensor that only depends explicitly on the electric ($\vec{E}$) and magnetic ($\vec{B}$) fields. \\


Since the analytic solution is found with the proper time as an independent variable, we need to solve its equation for a given $\delta t$, expressed as $\delta t = f(\delta \tau, \vec{E}, \vec{B})$, where $\vec{E}$ and $\vec{B}$ are functions of particle position in general. \\

With those considerations, the algorithm of pushing a particle can be outlined in the following three main steps:
\begin{enumerate}
    \item Project the initial condition of the particle into the two subspaces.
    \item Find the proper time step $\delta \tau$ for a given $\delta t$ through a Picard iteration procedure.
    \item Update the four-vector in the two subspaces according to the analytic solutions.
\end{enumerate}

\begin{figure*}
\includegraphics[width=12.5cm, height=15.5cm]{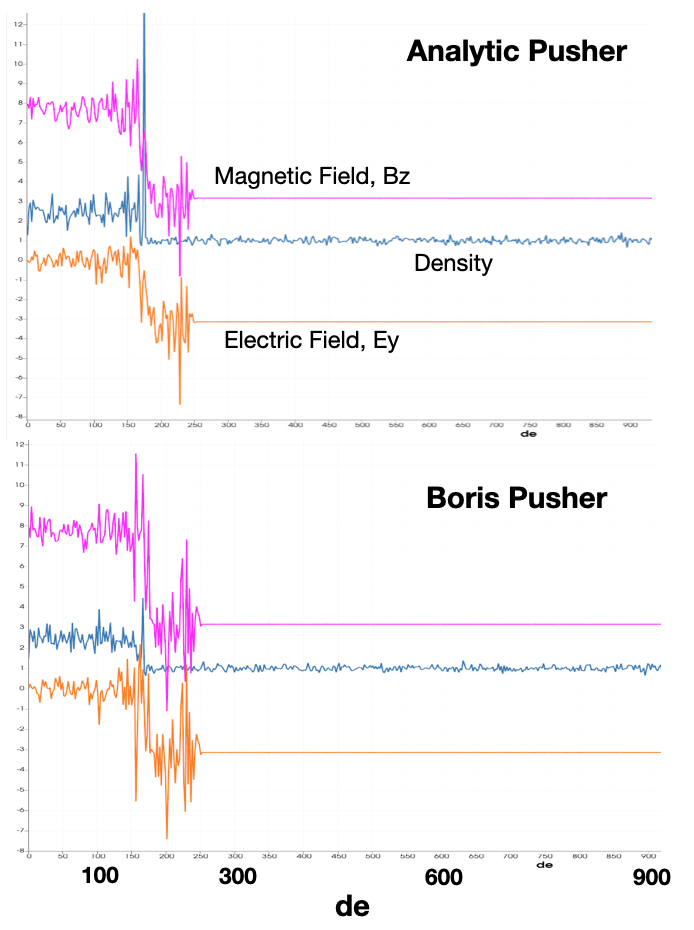} 
    \caption{Same simulation parameters as Figure 1, showing the precursor waves as a function of inertial length for the Analytic Pusher (top) and the Boris pusher (bottom). The panels shows the density (blue), magnetic field in the z direction (magenta) and electric field in the y direciton (orange).}
    \label{fig:apvbor1}
\end{figure*}

\begin{figure*}
\includegraphics[width=7.5cm, height=6.5cm]{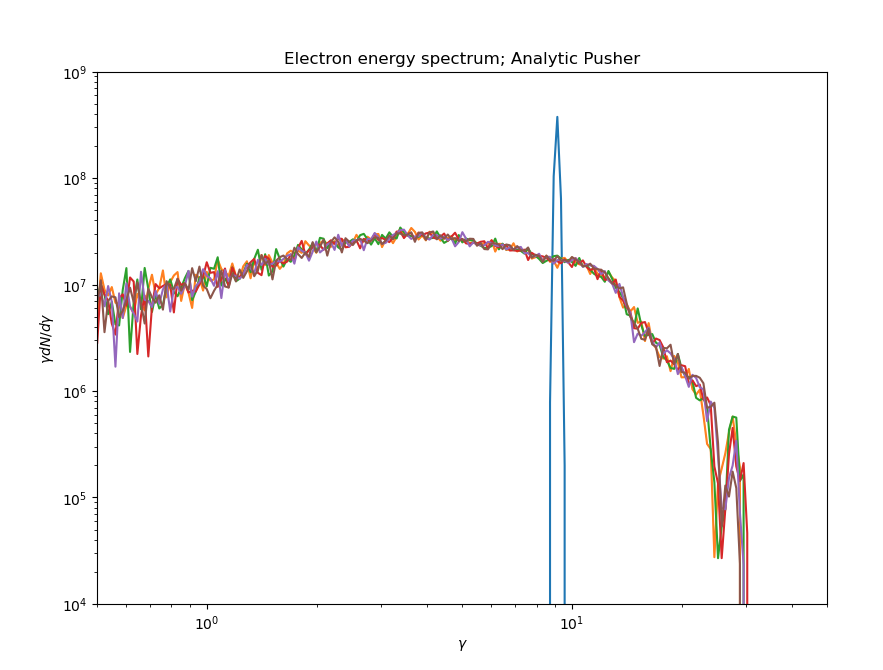}\includegraphics[width=7.8cm, height=6.4cm]{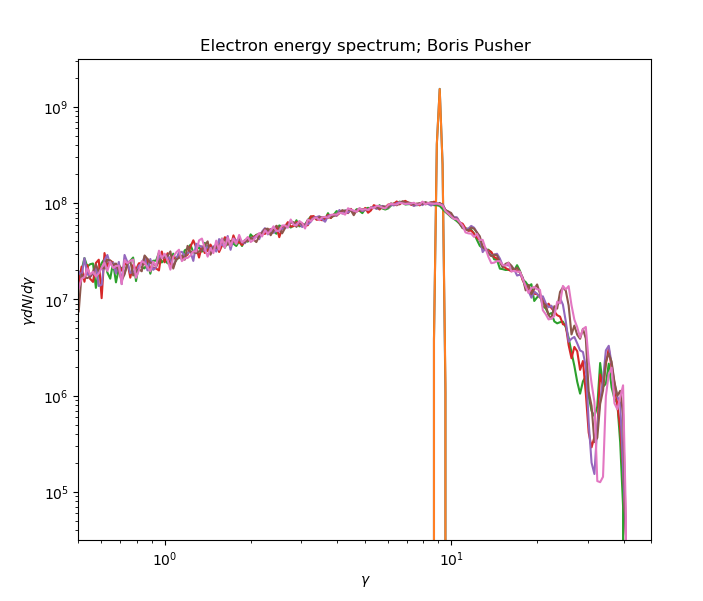}
    \caption{Same simulation parameters as Figure 1 but comparing the Analytic Pusher (bottom left) and to the Boris (bottom right) computing the spectrum over the region from x=65-70 de. Note that the analytic pusher has a much larger timestep with Lx/nx=4096/102400. The top panel shows the analytic pusher density (blue), magnetic field in the z direction (magenta) and electric field in the y direciton (orange).  As in Figure 1, the precursor wave is evident.}
    \label{fig:apvbor2}
\end{figure*}

Figures~\ref{fig:apvbor1} and ~\ref{fig:apvbor2} show the results from PIC runs implementing our analytic particle pusher. Figure~\ref{fig:apvbor1} is similar to Figure~\ref{fig:nbe}, showing a snapshot in time of the magnetic field, electric field, and particle density as a function of inertial length.  In the left panel of Figure~\ref{fig:apvbor2}, we present results of the simulation set-up similar to Figure~\ref{fig:elspecsig10} but employing our analytic particle pusher. For direct comparison, we present the results of the Boris pusher in the right panel. Note that the analytic pusher can be especially useful in cases where $r_{g}/\lambda_{D} \approx (B_{\rm PIC}\gamma_{e})^{-1/2} \ll 1$, when the gyro-radius is much smaller than the Debye length. We refer the reader to Chen et al. 2025 for a complete discussion of where the Boris pusher fails and how the analytic pusher can be particularly useful.\footnote{In particular, we find in that paper that the analytic pusher gets the ``right" answer (defined by the answer or data given for a highly resolved simulation) without having to resolve the gyro-radius or gyro-period.} In future work, we will apply this new technique to other astrophysical problems in which it can provide an advantage over the Boris pusher.

\section{Discussion and Conclusions}

Kinetic simulations of astrophysical plasmas are crucial to uncovering particle acceleration and emission mechanisms across the electromagnetic spectrum, from a wide range of astrophysical transients. They are especially important in extreme physical regimes, involving high magnetic fields or relativistic shocks. These simulations usually rely on standard particle-in-cell (PIC) techniques to capture the underlying physics of the plasmas, but simulations are expensive due to the need to resolve the smallest length and time scales in the problem.  In particular, for plasmas with high magnetic fields, the gyro-radius is sometimes prohibitively small, preventing modeling of the system over any observationally relevant timescales. \\

In this paper we have run a suite of simulations of relativistic one dimensional pair plasma shocks in a uniform background magnetic field oriented perpendicular to the direction of the flow.  We are motivated by previous works that have suggested such simulations may be relevant for the plasmas that produce fast radio burst transients.  We have followed the set-up of PS19, but covered a wider range of parameter space. \\

\noindent Our main results are as follows:
\begin{itemize}
    \item We reproduce the results of PS19 - namely, in a relativistic one dimensional shock with a background magnetic field there appears to be a sustained precursor wave that generates a population inversion of the pairs in a narrow region around the precursor and shock front, which can cause  synchrotron maser emission in such systems.  The continual flow of plasma through the shock allows for the population inversion to be steadily sustained.
     \item We find this inversion feature occurs across a range of magnetic field energy densities ($\sigma \approx 0.1 $--$ 200$) but begins to disappear at the highest magnetic fields ($\sigma > 250$).
     \item We also find the population inversion occurs for temperatures ranging from $T \approx 0.0001 $ -- $ .01 m_{e}c^{2}$, but begins to disappear at temperatures $T > 0.03 m_{e}c^{2}$.
     \item We find our newly developed analytic particle pusher, which relieves the constraint of having to resolve the gyro-radius in PIC simulations - gives similar results to the highly resolved simulations using the Boris pusher.
\end{itemize}

Our results have implications for models of coherent emission from FRBs. Our electron-positron plasma in one dimension is a reasonable representation of what is expected in regions around highly magnetized neutron stars \citep{HL06} \citep[and, again, the one dimensional simulation results are representative of what is found in similar two and three dimensional simulations][]{Bab20, Sir21}. In this magnetar scenario, a highly magnetized neutron star undergoes some type of a crust deformation that twists or rearranges the magnetic field, causing a magnetic pulse to propagate into the surrounding wind and generate a relativistic shock in the magnetosphere.  As mentioned in the introduction, the peak of the emission from a synchrotron maser has a weak dependence on the magnetization parameter and generally falls at a few times the plasma frequency.  FRBs are usually observed in the range of a few hundred MHz, which implies densities $n \sim 10^{7} cm^{-3}$, which is not unusual for the plasmas around pulsars or magnetars \citep{GJ69, Pot14}.\\

 We have demonstrated the utility of a novel analytic particle pusher for PIC simulations addressing astrophysical problems in which the gyro-radius is much smaller than the scale of the system and other plasma length scales (e.g. the Debye length and the inertial length).  This method has the potential to address a number of unsolved problems in extreme astrophysical environments - particularly those with high magnetic fields - and further our understanding of the physics in these systems.

\section*{Acknowledgements}
We thank Lorenzo Sironi for helpful comments, and acknowledge useful discussions with Federico Fraschetti and Manolis Drimalas. This work was supported by the U.~S. Department of Energy through Los Alamos National Laboratory (LANL).  LANL is operated by Triad National Security, LLC, for the National Nuclear Security Administration of U.S. Department of Energy (Contract No. 89233218CNA000001).   Research presented was supported by the Laboratory Directed Research and Development program of LANL project number 20230217ER. We acknowledge LANL Institutional Computing HPC Resources under project kinmhd. Additional research presented in this article was supported by the 
Laboratory Directed Research and Development program of Los Alamos National Laboratory under 
project number 20210808PRD1. LA-UR-25-22058

\section*{Data Availability}

 The data generated and used in this paper are available upon request.



\bibliographystyle{mnras}
\bibliography{ms} 








\bsp	
\label{lastpage}
\end{document}